\documentclass[10pt]{article}
\usepackage{graphicx}        
\usepackage{amssymb}
\usepackage{amsmath}
\usepackage{pdflscape}
\usepackage{tikz}
\usepackage[scriptsize]{subfigure}
\usepackage{amsmath}
\usepackage{soul}

\usepackage{float}

\usepackage[T1]{fontenc}
\usepackage{kpfonts,baskervald}

\usepackage[utf8]{inputenc}
\usepackage{authblk}

\usepackage{placeins}

\tikzset{cross/.style={cross out, draw=black, minimum size=2*(#1-\pgflinewidth), inner sep=0pt, outer sep=0pt},
cross/.default={2.5pt}}
\usepackage[margin=1in]{geometry}
\usepackage{parskip}

\DeclareGraphicsExtensions{.pdf,.png,.jpg,.eps} 
\usepackage{epstopdf}
\begin{document}

\title{Impact on power system transient stability of AC-line-emulation controllers of VSC-HVDC links}
\date{ }

\author{Javier Renedo, Lukas Sigrist, Luis Rouco, Aurelio Garcia-Cerrada}
\affil{Instituto de Investigaci\'on Tecnol\'ogica (IIT)\\ ETSI ICAI \\ Universidad Pontificia Comillas \\  Madrid, Spain \\ \{javier.renedo, lukas.sigrist, luis.rouco, aurelio\}@iit.comillas.edu}

\maketitle


\section*{Abstract}\label{sec:abstract}
High voltage direct current links based on voltage source converters~(VSC-HVDC) embedded in alternating current (AC) systems are receiving a great deal of attention recently because they can contribute positively to the flexibilisation of modern power systems. Among several possibilities, AC-line-emulation control has been highlighted as an simple-but-useful alternative for these type of systems. With this strategy, the power flow through the link is controlled proportionally to the angle difference between its two AC terminals and this provides self-adaptation of the power flow in case of contingencies in the parallel AC lines, naturally. Although this controller is mainly concerned with steady state, it can also have an impact on the dynamic behaviour of the system which has not been sufficiently analysed. Along this line, this paper analyses the impact of AC-line-emulation controllers of VSC-HVDC on power system transient stability. Nonlinear time-domain simulations were carried out by using PSS/E on a small test system with an embedded point-to-point VSC-HVDC link. The critical clearing time (CCT) of a test fault has been used to assess transient-stability margins of the whole system. The paper provides recommendations for the design of AC-line-emulation controllers in order to ensure that transient stability is not jeopardised.

\noindent \textbf{Index terms:} VSC-HVDC, AC-line emulation control, transient stability, power system stability

\noindent This is an unabridged draft of the following  paper (accepted for publication):
\begin{itemize}
	\item Javier Renedo, Lukas Sigrist, Luis Rouco and Aurelio Garcia-Cerrada, \emph{``Impact on power system transient stability of AC-line-emulation controllers of VSC-HVDC links"}, Proc. 14th IEEE/PES PowerTech Conference, Madrid, Spain, June 28th - July 2nd, 2021, pp. 1-6. 
\end{itemize}

\newpage

\section{Introduction}\label{sec:Intro}
\noindent High voltage direct current transmission with voltage source converters~(VSC-HVDC) is a key technology to increase transmission capacity of power systems and to facilitate the integration of renewable resources into the power system efficiently~\cite{Flourentzou2009}. Several VSC-HVDC installations are already in operation worldwide and many others are in a planning stage~\cite{Flourentzou2009,Inelfe2016}. The simplest way and, possibly the most popular one, of controlling a VSC-HVDC link embedded in and alternating current system (AC) is with constant power-flow although it is not the only alternative. For example,~\cite{Inelfe2016,Michi2019} proposed AC-line emulation with the HVDC-VSC link because transmission system operators (TSOs) are very familiar with the results of this strategy, among other reasons. AC-line emulation controls the power flow through an HVDC link proportionally to the difference between the voltage angles of the two AC terminals of the link. This is exactly what would happen if an AC line were in place. Such a controller has a very simple structure consisting of a proportional gain, which is the inverse of the reactance of the emulated AC line and a low-pass filter. However, it has a natural capability of self-adaptation of the power flow to system contingencies. For example, if an AC line in parallel with the link trips, the VSC-HVDC link will naturally, and automatically, change its power flow. Examples of VSC-HVDC installations in operation with AC-line-emulation control are Mackinac back-to-back interconnector (USA)~\cite{Marz2014,Danielsson2015}, INELFE Spain-France interconnector through the Catalonian Pyrenees~\cite{Inelfe2016} and Piedmont-Savoy France-Italy interconnector~\cite{Michi2019,Michi2020}, which connects Grande Ile substation (France) with Piossasco substation (Italy).

Although the AC-line-emulation control of a VSC-HVDC link is a steady-state controller, it will have an impact on the dynamic behaviour of the system, which is a major concern for TSOs. For example,~\cite{IIT_ACemul2019,Inelfe2020,Michi2019,Michi2020} showed that an AC-line-emulation control of a VSC-HVDC link can have a negative impact on inter-area oscillations for certain values of the time constant of the low-pass filter within the controller. In fact,  Spanish and French TSOs changed the value of that time constant from its initial value ($T=0.75$~s) to a much slower one ($T=50$~s)  in the INELFE interconnector, in order to avoid this problem~\cite{Inelfe2020}.

Power system transient stability, also known as rotor-angle stability under large distrubances~\cite{Kundur2004}, is a critical limiting factor in heavily loaded transmission systems with long AC lines and several publications have proposed supplementary controllers in VSC-HVDC systems to improve transient stability~\cite{Johansson2004,Latorre2008,Lukas2015,Machowski2013,Eriksson2014a,Fuchs2014,Tang2016,iitcontrolQ2017,Fan2018,iitcontrolQ_local2019,JuanCarlos_transient_stab2020}. Since the supplementary controllers proposed in~\cite{Johansson2004,JuanCarlos_transient_stab2020} include terms depending on angle differences with good results, one may expect that AC-line-emulation controllers in VSC-HVDC links could also have a beneficial effect on transient stability.  However, there are several aspects worth investigating further, such as: (a) the effect of the controller parameters on the operating point; (b) the effect of the gain of the controller on transients and (c) the influence of the low-pass filter time constant. 

This paper analyses the impact of AC-line-emulation controllers of VSC-HVDC on transient stability. To the best of the authors' knowledge, such study has not been published before. The specific contributions of this paper are as follows:
\begin{itemize}
	\item A detailed study on the impact of AC-line-emulation controllers of VSC-HVDC on transient stability. 
	\item Design recommendations for AC-line-emulation controllers in order to ensure that transient stability is not jeopardised.
\end{itemize} 


\FloatBarrier

\section{VSC-HVDC systems}\label{sec:vsc_hvdc_systems}
\begin{figure}[!htbp]
\begin{center}
\includegraphics[width=0.8\columnwidth]{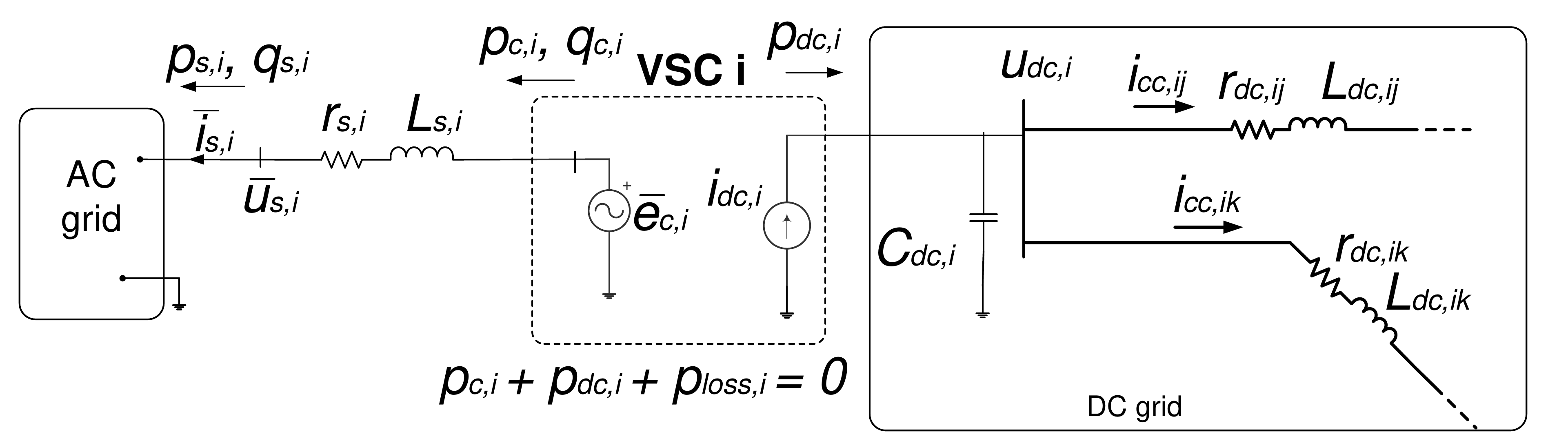}
\caption{VSC and the DC grid based on ~\protect\cite{Cole2010,Beerten2014}.}
\label{fig:vsc_dcgrid_model}
\end{center}
\end{figure}

\noindent A point-to-point VSC-HVDC system consists of two VSC stations connected through a DC line. Fig.~\ref{fig:vsc_dcgrid_model} shows the fundamentals of the dynamic model of a VSC station connected to an AC grid and to a generic DC grid (i.e. a multi-terminal VSC-HVDC system), following the guidelines of~\cite{Cole2010, Beerten2014}. 

Each VSC has an outer controller and an inner current controller. With the outer controller, each VSC can control (a) the active-power injection~($p_{s,i}$) or the DC voltage~($u_{dc,i}$) (by manipulating the setpoint of the $d$-axis current), and (b) the reactive-power injection~($q_{s,i}$) or the magnitude of the AC voltage~($u_{s,i}$) (by manipulating the setpoint of the $q$-axis current). The setpoints of $dq$-axes currents are the inputs of the inner controller. The model used in this paper was intended at studying electromechanical transients, where the slow electromechanical dynamics of the power system are of interest. Hence, the outer controllers were modelled in detail, while the closed-loop system of the inner controllers were represented by first-order transfer functions. The models of the VSC stations were implemented with their operating limits (P/Q limits, current limit and maximum modulation index). At the DC side, VSC stations are represented as current sources ($i_{dc,i}$). The equivalent capacitance~($C_{dc,i}$) of each DC bus aggregates the equivalent capacitor of the VSC plus the contribution of the capacitances of DC lines connected the DC bus.  DC lines are modelled with series reactances~($r_{dc,ij}$) and a series inductances~($L_{dc,ij}$). All the details of the implementation of the full model can be found in~\cite{jrenedoPSSE2017}.

Active-power injections into  AC ($p_{c,i}$) and DC sides ( $p_{dc,i}$) of a VSC are related by:
\begin{equation}\label{eq:VSC_energy_conservation}
p_{c,i} + p_{loss,i}+ p_{dc,i}=0
\end{equation}
where $p_{loss,i}$ are the  losses of the VSC, calculated as~\cite{Daelemans2009}:
\begin{equation}\label{eq:VSC_ploss}
p_{loss,i} = a_i + b_i \cdot i_{s,i} + c_i \cdot i_{s,i}^2
\end{equation}
and $a_{i}$, $b_{i}$ and $c_{i}$ are specific parameters for each VSC. 

The initial operating point of the hybrid AC/DC power system must be calculated using a power-flow algorithm, for example, the work reported here used the sequential one reported in~\cite{Beerten2012}.

\FloatBarrier
\section{AC-line-emulation controllers for HVDC-VSC links}\label{sec:vsc_hvdc_ACemul}
\begin{figure}[!htbp]
\begin{center}
\includegraphics[width=0.8\columnwidth]{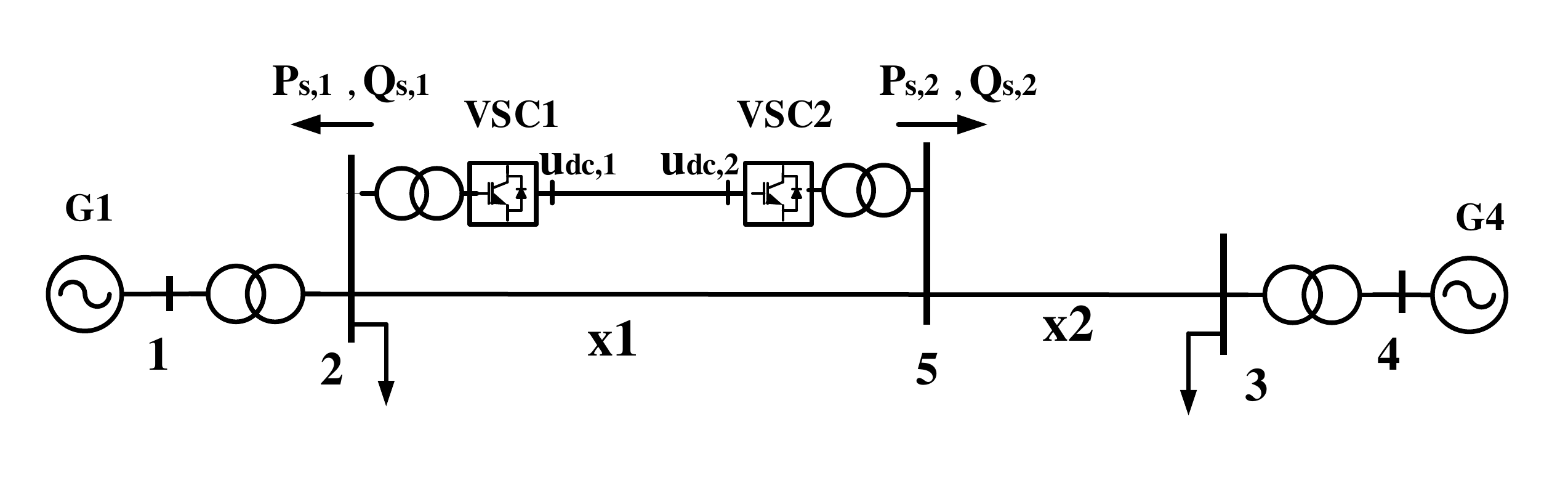}
\caption{AC system with an embedded VSC-HVDC link.}
\label{fig:vsc_hvdc_2bus_system}
\end{center}
\end{figure}

\noindent Fig.~\ref{fig:vsc_hvdc_2bus_system} shows a VSC-HVDC link embedded in an AC system. Let's assume that VSC1 controls its power injection into the AC side ($p_{s,1}$) and VSC2 controls its DC-voltage ($u_{dc,2}$). An AC-line-emulation controller for the VSC-HVDC link would manipulate the active-power set point as~\cite{Inelfe2016}:
\begin{equation}\label{eq:VSC_HVDC_ACline_emul}
p_{s,1}^{ref} = p_{s,1}^{cons} - \frac{K}{1+T s} (\delta_{s,1} - \delta_{s,2})
\end{equation}
where $p_{s,1}^{cons}$ is a constant setpoint (pu), $K$ is the controller gain (pu/rad), $T$ is the time constant of a low-pass filter (s) and $\delta_{s,i}$ in rad (with $i=1,2$) is the angle measured at the AC connection point $s,i$, which can be obtained by means of Phasor Measurement Units~(PMUs), for example. The controller gain is calculated as the inverse of the reactance of the emulated AC line ($x_{HVDC}$):
\begin{equation}\label{eq:VSC_HVDC_ACline_emul_xhvdc}
K = \frac{1}{x_{HVDC}}
\end{equation}
When a disturbance occurs, power flows through the VSC-HVDC link and through its parallel line (2-5) (see Fig.~\ref{fig:vsc_hvdc_2bus_system}) will be distributed, in steady state, according to the emulated reactance, $x_{HVDC}$, and the impedance of parallel AC lines ($x_1$ in the small test system of Fig.~\ref{fig:vsc_hvdc_2bus_system}).  Therefore, $K$ should be designed according to steady-state preferences of the TSO. Parameter $T$ should be designed to avoid negative interactions with the dynamic behaviour of the system.  Notice that, since the AC-line-emulation controller is mainly concerned with the steady-state result, the bandwidth of the low-pass filter is not a critical parameter~\cite{Inelfe2020}.

The behaviour of an AC-line-emulation controller on transient stability has been compared with the one obtained in the case in which the VSC-HVDC link has a constant active-power set point, but with the same initial operating point in both cases. For this purpose, it was useful writing (\ref{eq:VSC_HVDC_ACline_emul}) as:
\begin{equation}\label{eq:VSC_HVDC_ACline_emul2}
p_{s,1}^{ref} = p_{s,1}^{ini} - \frac{K}{1+T s} (\Delta \delta_{s,1} - \Delta \delta_{s,2})
\end{equation}
where $p_{s,1}^{ini}=p_{s,1}^{cons}-K(\delta_{s,1}^{0} - \delta_{s,2}^{0})$ is the initial power injection of the operating point (obtained by power-flow calculation), $\delta_{s,i}^{0}$ is the initial angle and $\Delta \delta_{s,i}=\delta_{s,i}-\delta_{s,i}^{0}$ is the angle increment during the transients.

\FloatBarrier
\section{Results}\label{sec:results}
\noindent Simulations were carried out with Kundur's two-area test system with an embedded point-to-point VSC-HVDC link of 1000~MVA, as shown in Fig.~\ref{fig:Kundur_vsc_hvdc_pap}. Data are provided in the Appendix. Simulations were carried out in PSS/E, with the model presented in~\cite{jrenedoPSSE2017}. The AC-line-emulation controller of VSC1 was implemented as a user-defined supplementary controller. Converter VSC1 (connected to bus 6) controls its active-power injection into the AC grid ($p_{s,1}$), while converter VSC2 (connected to bus 10) controls its DC voltage. Both converters control their reactive-power injections ($q_{s,i}$) down to zero. Since the power flow of the VSC-HVDC link will go from the left-hand side area to the right-hand side one (Fig.~\ref{fig:Kundur_vsc_hvdc_pap}). Notice that   $P_{HVDC}=-P_{s,1}$, for the sake of clarity.

\begin{figure}[!htbp]
\begin{center}
\includegraphics[width=0.8\columnwidth]{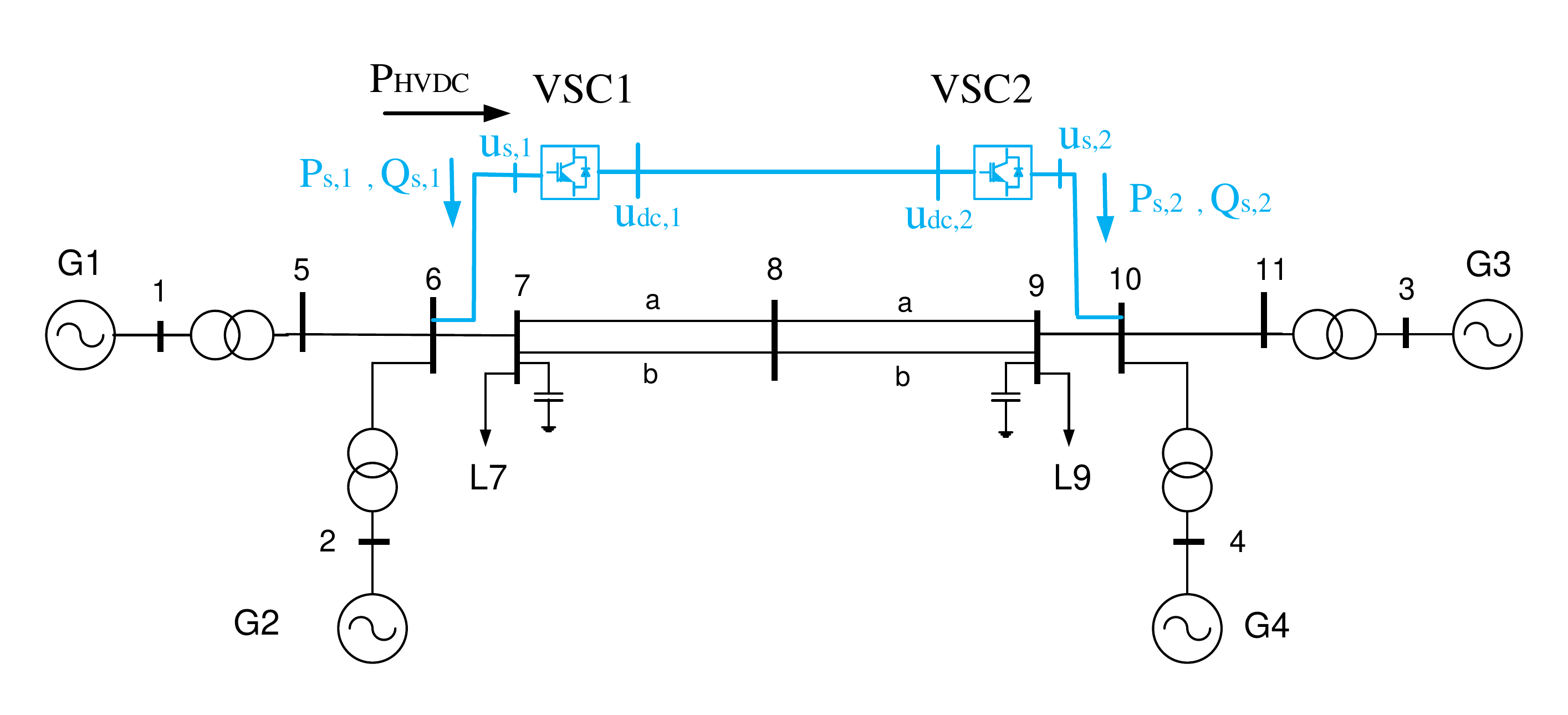}
\caption{Kundur's two-area system with an embedded VSC-HVDC link.}
\label{fig:Kundur_vsc_hvdc_pap}
\end{center}
\end{figure}

The value of the gain of the AC-line-emulation controller was chosen close to the value of the reactance of the parallel AC path, which has a total reactance, written in pu of system's bases (100 MVA and 220 kV), of:
\begin{eqnarray}\label{eq:Kundur_AC_lines_impedance}
x_{AC,eq} &=& x_{67} + x_{78a}//x_{78b} + x_{89a}//x_{89b} + x_{9,10} \\ \nonumber
&=& 0.025 + 0.11/2 + 0.11/2 + 0.025 = 0.1150 \mbox{ pu}
\end{eqnarray}
More precisely, the equivalent reactance of the emulated AC line of the VSC-HVDC link was set to $x_{HVDC}=0.1$~pu, i.e. $K=1/x_{HVDC}=10$~pu/rad (in pu of system bases).  However, in the controller, the gain is typically written in pu of the converter rating (1000 MVA and 220 kV) giving $K=10 \cdot 100/1000= 1$~pu/rad.

Several values for the  filter time constant of the AC-line-emulation controller were studied with especial attention to those used in the France-Spain INELFE VSC-HVDC system (initially $T=0.75$~s and updated to $T=50$~s)~\cite{Inelfe2020}. 

The following cases will be compared:
\begin{itemize}
	\item Constant P: $P_{HVDC}^{cons}=P_{HVDC}^{ini}=438$~MW (This value is equal to the one obtained in steady state when using the AC-line-emulation controller in the VSC-HVDC link).
	\item AC-line emulation: $P_{HVDC}^{cons}=0$, $K=1$~pu/rad  and $T=0.75$~s. The initial power injection obtained by power-flow calculation is: $P_{HVDC}^{ini}=438$~MW.
	\item AC-line emulation: $P_{HVDC}^{cons}=0$, $K=1$~pu/rad (nominal pu) and $T=50$~s. The initial power injection obtained by power-flow calculation is: $P_{HVDC}^{ini}=438$~MW.
\end{itemize}

\FloatBarrier

\subsection{Performance of AC-line emulation controller}\label{sec:results_ACemul_performance}
\begin{figure}[!htbp]
\begin{center}
\includegraphics[width=0.60\columnwidth]{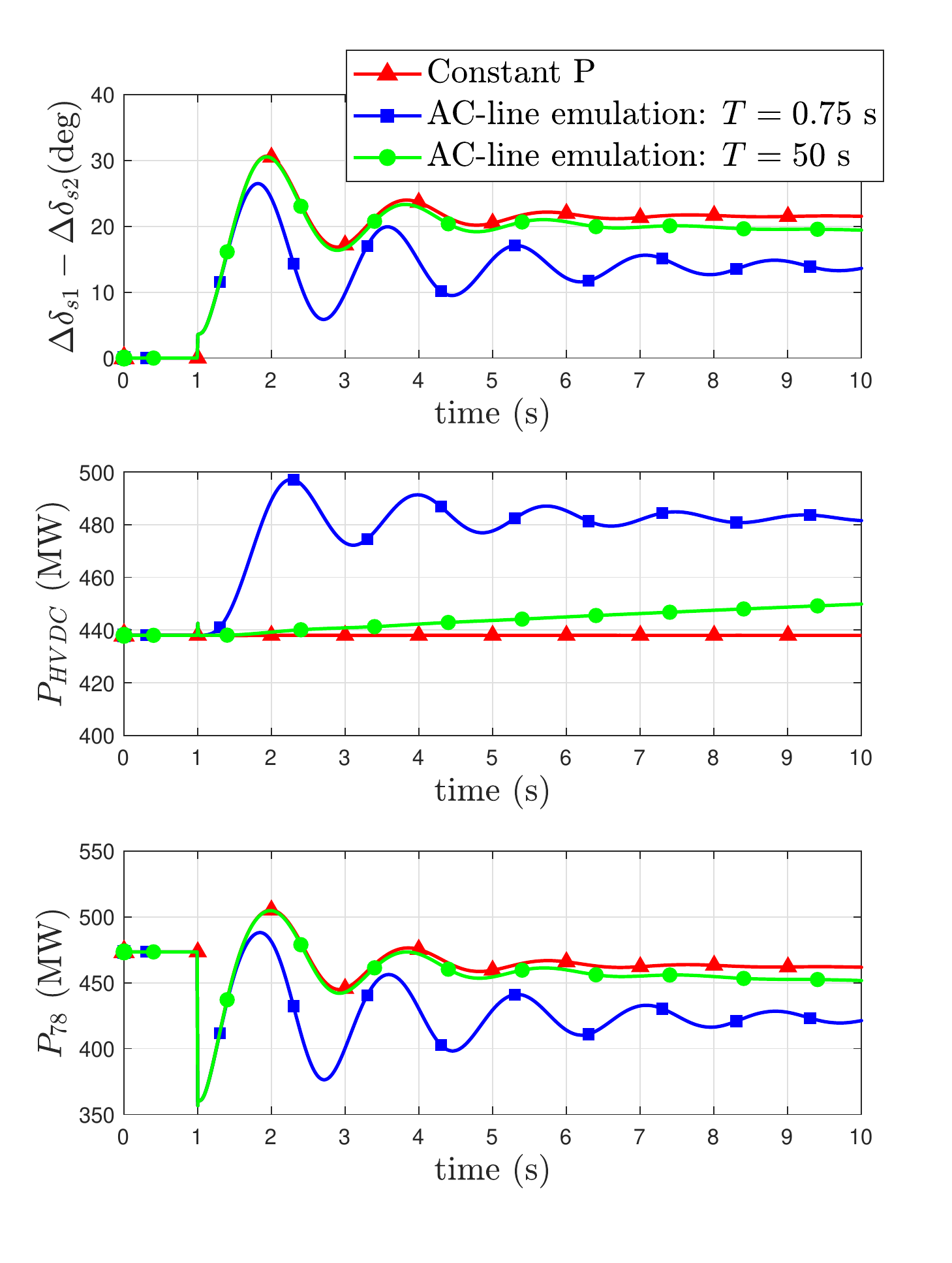}
\caption{Increment of angle difference ($\Delta \delta_{s,1}-\Delta \delta_{s,2}$), active power flow through the VSC-HVDC ($P_{HVDC}$) and through the parallel AC corridor ($P_{78}$). Gain of the AC-line-emulation controller: $K=1$~pu/rad.}
\label{fig:Kundur_sim0_A2_Phvdc}
\end{center}
\end{figure}

\begin{figure}[!htbp]
\begin{center}
\includegraphics[width=0.60\columnwidth]{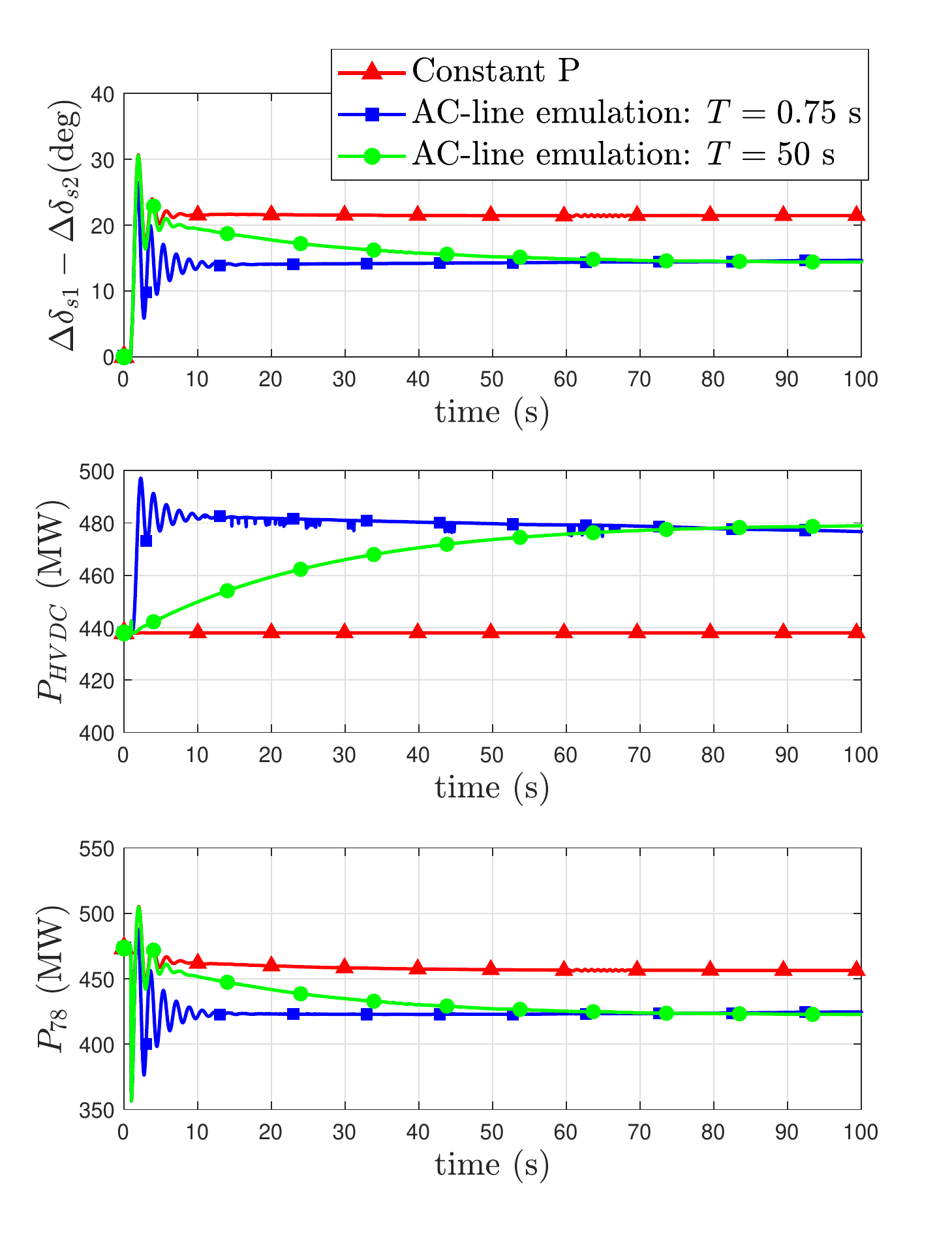}
\caption{Increment of angle difference ($\Delta \delta_{s,1}-\Delta \delta_{s,2}$), active power flow through the VSC-HVDC ($P_{HVDC}$) and through the parallel AC corridor ($P_{78}$). Gain of the AC-line-emulation controller: $K=1$~pu/rad. Sim. time: $t_{final}=100$~s.}
\label{fig:Kundur_sim0_A2_Phvdc_tsim_100s}
\end{center}
\end{figure}

\noindent First of all, the performance of the AC-line-emulation controller is illustrated by simulating the disconnection of circuit \emph{a} of line 7-8 (see Fig.~\ref{fig:Kundur_vsc_hvdc_pap}), which is initially carrying 236.40~MW in the initial operating point. Fig.~\ref{fig:Kundur_sim0_A2_Phvdc} shows the increment of the angle difference between the two AC terminals of the VSC-HVDC link ($(\Delta \delta_{s,1} - \Delta \delta_{s,2}$), the power flow of the VSC-HVDC link ($P_{HVDC}$) and the total power flow of the AC line 7-8 (through both circuits) ($P_{78}$).  Fig.~\ref{fig:Kundur_sim0_A2_Phvdc_tsim_100s} shows the same variables in a longer time span (100 s).

The disconnection of line 7-8\emph{a} provokes an increment on the angle difference between the terminals of the VSC-HVDC link. In the case of constant P (i.e. the VSC-HVDC maintains the power flow through the DC link constant) only the power flow through line 7-8\emph{a} increases. With an AC-line-emulation controller with $T=0.75$~s, both, the VSC-HVDC and the parallel AC corridor increase their power flow rapidly and the total increment of the power flow is distributed between the VSC-HVDC link and the AC corridor according to the equivalent reactance of the emulated AC line of the VSC-HVDC and the equivalent impedance of the AC corridor. With an AC-line-emulation controller with $T=50$~s, the same steady-state solution than with $T=0.75$~s is reached, eventually, but the transient is much slower~(Fig.~\ref{fig:Kundur_sim0_A2_Phvdc_tsim_100s}). 

\FloatBarrier
\vspace{-2cm}
\subsection{Fault simulation}\label{sec:results_fault}
\begin{figure}[!htbp]
\begin{center}
\includegraphics[width=0.60\columnwidth]{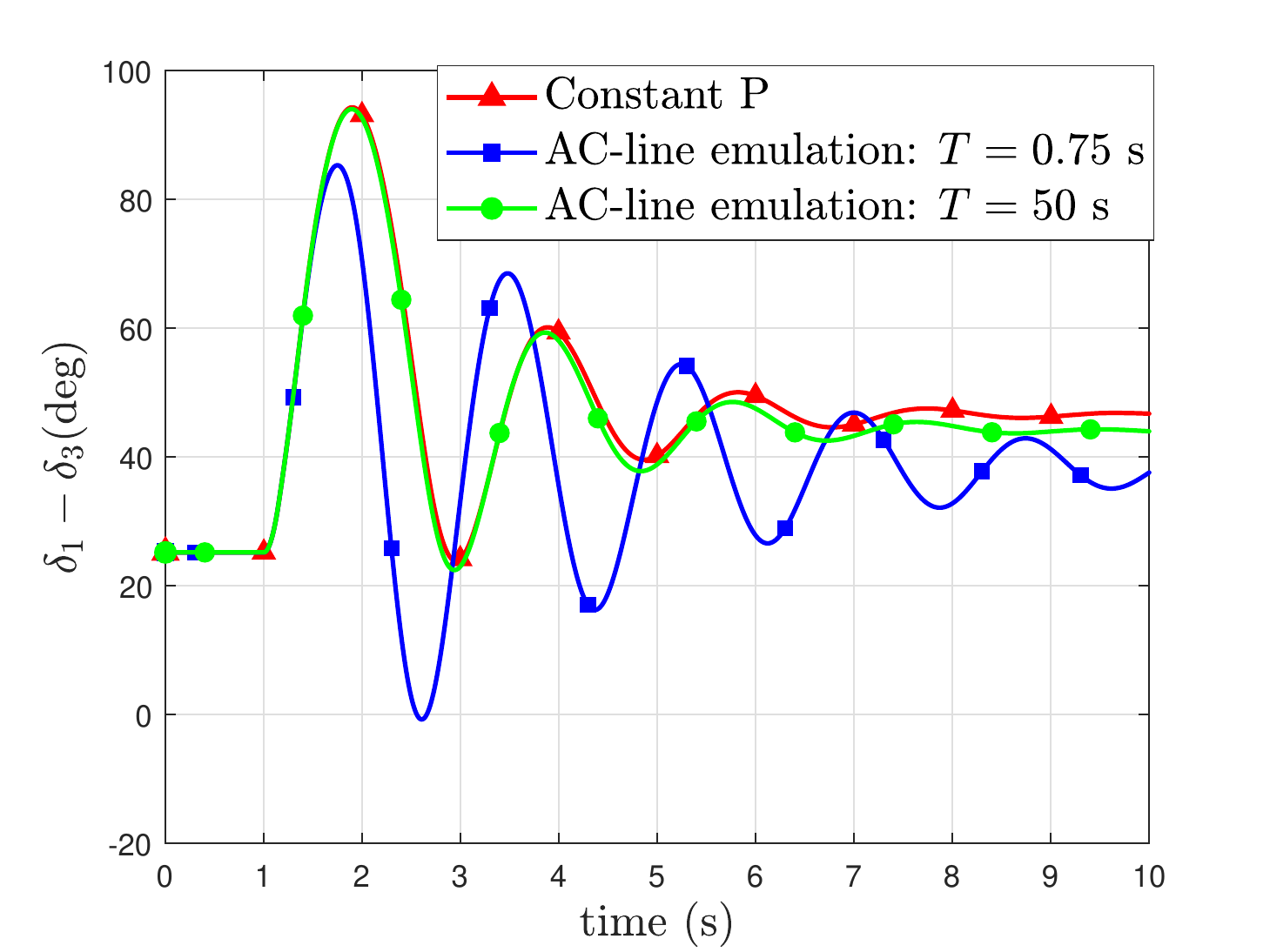}
\caption{Generator-angle difference. Gain of the AC-line-emulation controller: $K=1$~pu/rad.}
\label{fig:Kundur_sim1_A2_Angles}
\end{center}
\end{figure}

\begin{figure}[!htbp]
\begin{center}
\includegraphics[width=0.60\columnwidth]{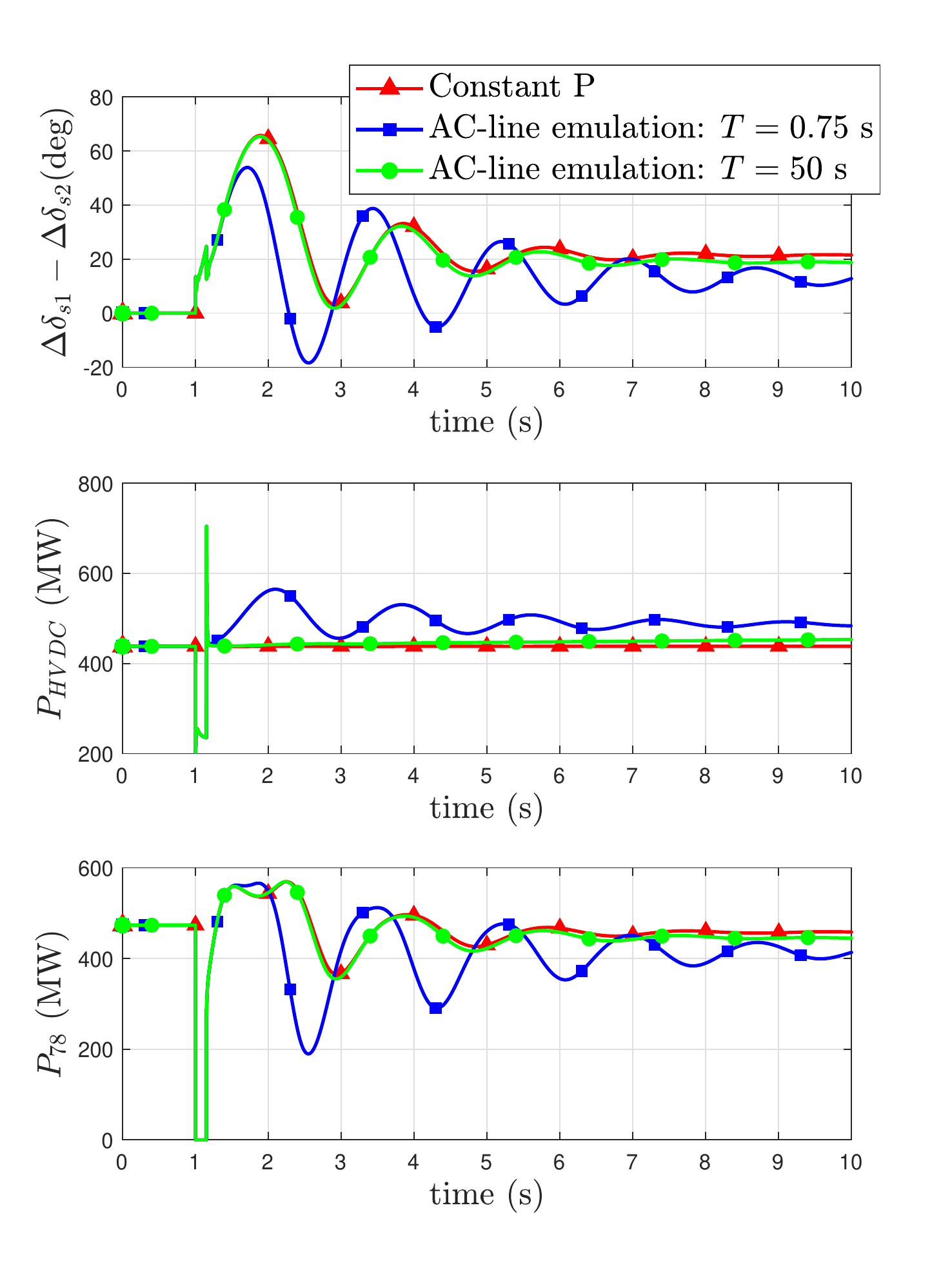}
\caption{Increment of angle difference ($\Delta \delta_{s,1}-\Delta \delta_{s,2}$), active power flow through the VSC-HVDC ($P_{HVDC}$) and through the parallel AC corridor ($P_{78}$). Gain of the AC-line-emulation controller: $K=1$~pu/rad.}
\label{fig:Kundur_sim1_A2_Phvdc}
\end{center}
\end{figure}

\noindent A three-phase-to-ground short circuit was applied to line 7-8\emph{a} (close to bus 7) at $t=1$~s. The fault was cleared after 150 ms by disconnecting the circuit. Fig.~\ref{fig:Kundur_sim1_A2_Angles} shows the voltage angle difference between the buses of generators G1 and G3~(see Fig.~\ref{fig:Kundur_vsc_hvdc_pap}). Fig.~\ref{fig:Kundur_sim1_A2_Phvdc} shows the results associated to the VSC-HVDC link. Only a time span of 10~s has been used, since this is the time frame of interest for transient stability.  The case with AC-line-emulation controller with $T=0.75$~s seems to improve transient stability with respect to the case of constant P and the angle difference in the first swing is smaller (Fig.~\ref{fig:Kundur_sim1_A2_Angles}). However, the former reduces the damping ratio of the inter-area mode with respect to the latter and the angle difference oscillates over a longer period of time. This is consistent with the findings of~\cite{IIT_ACemul2019,Inelfe2020,Michi2019,Michi2020}. Notice that, in general, the overall transient stability of the system could depend on the behaviour of the system during the first swing, but it can also be affected indirectly by poorly damped electromechanical oscillations, although they are different phenomena. For example, poorly damped electromechanical oscillations could facilitate loss of synchronism in the first swing for faults with longer clearing times. As expected, the behaviour of the cases with AC-line-emulation controller with a large filter time constant ($T=50$~s) is very similar to the result with constant P because the former responds very slowly and has no significant effect.

\FloatBarrier

\subsection{Critical clearing times (CCTs)}\label{sec:results_CCT}
The critical clearing time (CCT) of a certain fault is defined as the maximum clearing time that the system can withstand without provoking loss of synchronism. The CCT is normally used as an indicator of the transient-stability margin of a power system. The CCT of the fault simulated (short circuit at line 7-8\emph{a} (close to bus 7) cleared by disconnecting the circuit) for different cases will be compared in this section. The cases are as follows:
\begin{itemize}
	\item Case A:
\begin{itemize}
	\item Constant P: $P_{HVDC}^{cons}=P_{HVDC}^{ini}=438$~MW.
	\item AC-line emulation (AC-LE, for short): $P_{HVDC}^{cons}=0$, $K=1$~pu/rad. Initial power flow: $P_{HVDC}^{ini}=438$~MW.
\end{itemize}
	\item Case B:
\begin{itemize}
	\item Constant P: $P_{HVDC}^{cons}=P_{HVDC}^{ini}=556.30$~MW.
	\item AC-LE: $P_{HVDC}^{cons}=0$, $K=2$~pu/rad. Initial power flow: $P_{HVDC}^{ini}=556.30$~MW.
\end{itemize}
	\item Case C:
\begin{itemize}
	\item Constant P: $P_{HVDC}^{cons}=P_{HVDC}^{ini}=645.30$~MW.
	\item AC-LE: $P_{HVDC}^{cons}=0$, $K=4$~pu/rad. Initial power flow: $P_{HVDC}^{ini}=645.30$~MW.
\end{itemize}
\end{itemize}
Notice that, for each case, the initial operating point  with AC-line-emulation control was obtained by a power-flow study and this value was also used in the case of constant P, for comparison purposes. To start with, time constant of the low-pass filter was was set to $T=0.75$~s (i.e. the value initially proposed for this filter). Recall that $K$ always is written in pu's with respect to the converter nominal power.

CCTs obtained for each case are shown in Table~\ref{tab.Kundur_mach_CCT_Fault_I}. Results show that the AC-line-emulation controller could jeopardise transient stability for certain values of $T$.  For example, in Case C, the CCT is reduced from 387~ms (constant P) to 118~ms when using AC-line-emulation control (for values $T=0.75$~s). 

\begin{table}[!htbp]
\begin{center}
\caption{Fault critical clearing times (CCT).}
\begin{tabular}{|l|ccc|}
\hline
  CCT (ms) & Case A&  Case B  & Case C \\ 
 & ($K=1$~pu/rad) & ($K=2$~pu/rad)   & ($K=4$~pu/rad)  \\ 
\hline 
Constant P& 359 &298 & 370   \\
AC-LE ($T=0.75$~s) & 325& 210 &  118   \\  
 \hline
\end{tabular}
\label{tab.Kundur_mach_CCT_Fault_I}
\end{center}
\end{table}

\begin{figure}[!htbp]
\begin{center}
\includegraphics[width=0.7\columnwidth]{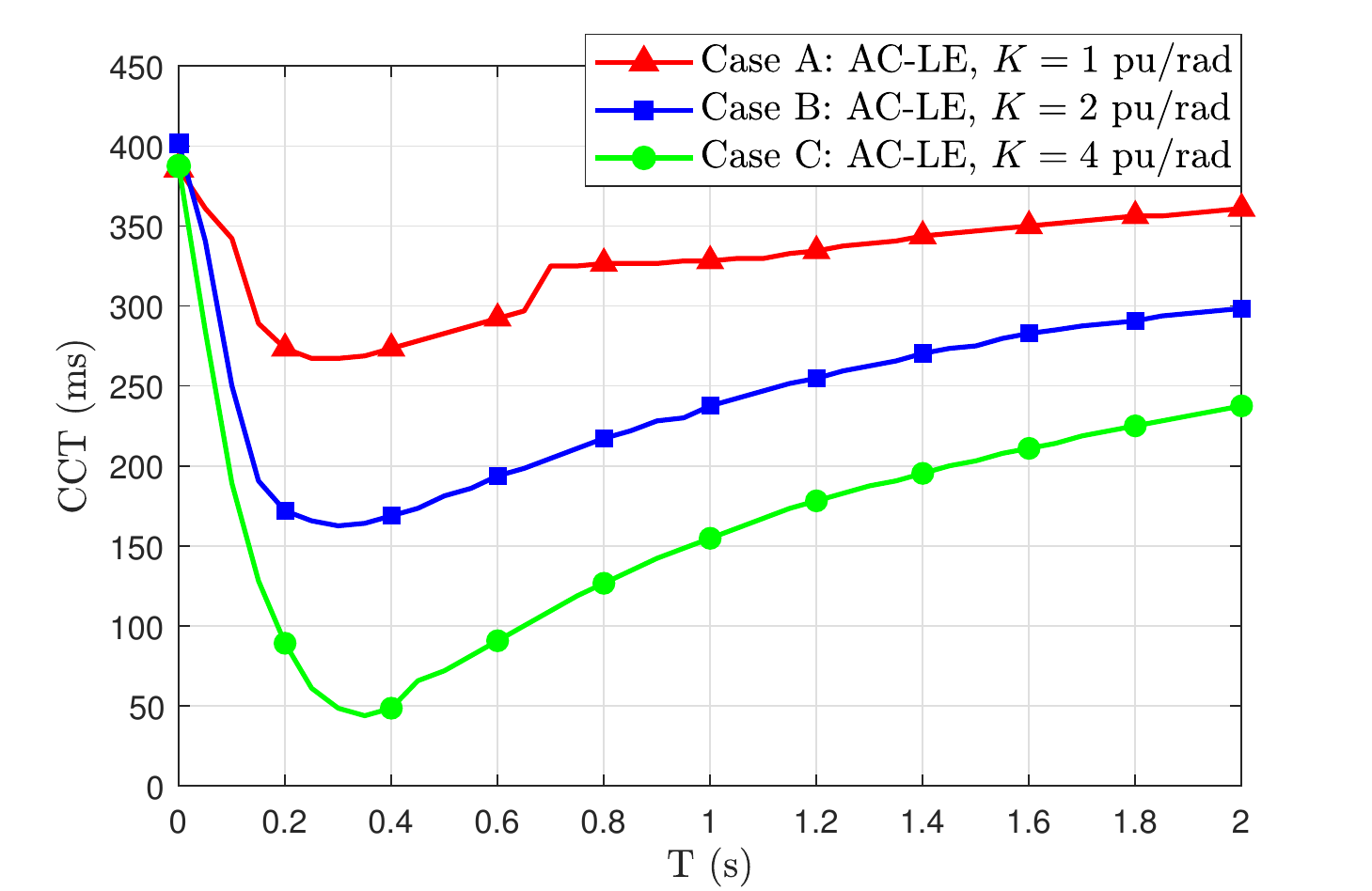}
\caption{Fault CCTs. $T$ is changed from 0 to 2 s (steps of 0.05~s). }
\label{fig:Kundur_CCTs_T_short}

\includegraphics[width=0.7\columnwidth]{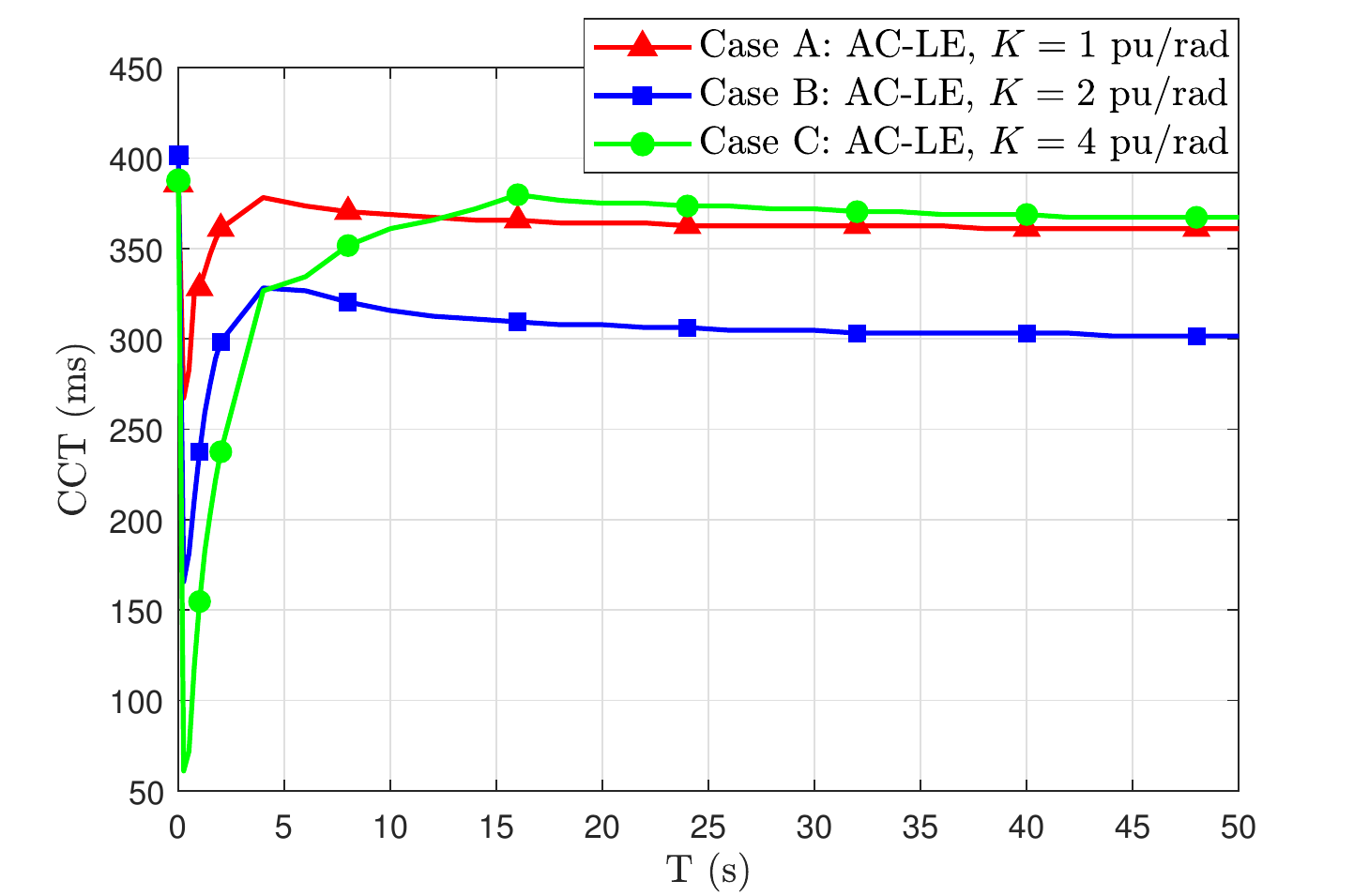}
\caption{Fault CCTs. $T$ is changed from 0 to 50 s (steps of 5~s). }
\label{fig:Kundur_CCTs_T_long}
\end{center}
\end{figure}

A more in-depth analysis is carried out now, in order to understand the pattern followed by CCTs when $K$ and $T$ change. Fig.~\ref{fig:Kundur_CCTs_T_short} shows fault CCTs obtained for different cases when $T$ is changed from 0 to 2 s (with steps of 0.05 s) while Fig.~\ref{fig:Kundur_CCTs_T_long} shows fault CCTs for values of $T$ between 0 and 50 s (with steps of 5 s). Results show that very low values of $T$ (almost zero) produce high CCTs. However, intermediate values of $T$ decrease the CCTs. Lower values of the CCTs in these intermediate values of $T$ with high values of $K$. For slower values of $T$, the CCTs recover. Final values of the CCTs, obtained for extremely high values of $T$, are the same as the ones obtained with constant P. As a general conclusion, an AC-line-emulation controller should be either very fast (extremely low values of $T$) or very slow (extremely high values of $T$) in order to avoid jeopardising transient stability. Since, previous work concluded that AC-line-emulation controllers could also have a negative impact on inter-area-oscillation damping for intermediate values of $T$~\cite{IIT_ACemul2019,Inelfe2020,Michi2019,Michi2020}, seeking a large value of $T$ is a reasonable and save strategy (e.g. $T\geq 20\;{\rm s}$).

\FloatBarrier
\section{Conclusions}\label{sec:Conclusions}
\noindent This paper has explored the performance of AC-line-emulation controllers in VSC-HVDC links. The following conclusions can be drawn from the results obtained. 
\begin{itemize}
	\item AC-line-emulation controllers in VSC-HVDC links can have either positive or negative impact on transient stability, depending on the values of the time constant of the low-pass filter, $T$.
	\item Summarising, one can conclude that AC-line-emulation controllers can be beneficial for transient stability if the controller is very fast (extremely low values of $T$) or if the controlller is very slow (high values of $T$). Intermediate values of $T$ can jeopardise transient stability.
	\item Since AC-line-emulation controllers are steady-state-orientated controllers and taking into account that intermediate values of $T$ are dangerous for transient stability (as shown in this work) and  for inter-area-oscillation damping (as shown in previous work), large values of $T$ are recommended ($T \geq 20\;{\rm s}$).
\end{itemize}

\FloatBarrier
\section*{Appendix: Test system data}\label{app:case_study1}

\noindent Data of the original two-area Kudur's test system can be found in~\cite{Kundur1994}. Nominal voltage of the transmission grid and the nominal frequency (230~kV and 60~Hz, respectively) were changed to n 220~kV and 50~Hz in this study. Besides, a critical case for transient stability was achieved using:
\begin{itemize}
	\item Inertia constants: $H_1=H_2=4.5$~s, $H_3=H_4=4.175$~s.
	\item Loads: L7: 467MW, 100MVAr;  L9: 2267MW, 0MVAr.
\end{itemize}

Parameter values used for the dynamic models of synchronous machines, governors, excitation systems and power system stabilisers (PSSs) are detailed in~\cite{iit_coord_control_delay2019}. In the dynamic simulations, loads were modelled with a constant current in the active-power part and with a constant impedance in the reactive-power part.

Parameters of the VSC-HVDC link are written in Table~\ref{tab:sim1_system_parameters}.

\FloatBarrier


\begin{table}[H]
\begin{center}
\caption{Parameters of the VSC-HVDC link. VSC's rating are base values for p.u.} 
\begin{tabular}{l|c}
\hline
VSC rating, DC voltage, AC voltage & 1000 MVA, $\pm 320$ kV, $300$ kV   \\
Configuration & Symmetrical monopole \\
Max. active (reactive) power & $\pm 1000$ MW ($\pm 450$ MVAr) \\
Max. current & 1 p.u ($d$-axis priority)  \\
Max. DC voltage & $\pm 10$ \% \\
Max. modulation index &1.31 p.u  \\ 
Current-controller time constant ($\tau$) & 5 ms  \\
Connection resistance ($r_s$)/reactance ($x_s$) & 0.02 p.u / 0.20 p.u \\
(reactor + 300/220 kV transformer) &  \\
P prop./int. control:   ($K_{d,p1}$/$K_{d,i1}$) & 0/0 \space \space (i.e. $i_{d,i}^{ref}=p_{s,i}^{ref}/u_{s,i}$)\\
Vdc prop./int. control  ($K_{d,p2}$/$K_{d,i2}$) & 10 p.u/20 p.u/s \\
Q-control prop./int. control: ($K_{q,p1}$/$K_{q,i1}$) & 0/0 \space \space (i.e. $i_{q,i}^{ref}=-q_{s,i}^{ref}/u_{s,i}$)  \\
VSCs' loss coefficients ($a$/$b$) in p.u. & 5.25/1.65 $\times 10^{-3}$ p.u. \\
VSCs' loss coefficients ($c_{rec}$/$c_{inv}$) in p.u. & 2.10/3.14 $\times 10^{-3}$ p.u. \\
Eq. DC-capacitance of each VSC ($C_{VSC,i}$) & 193.57 $\mu$F  \\ \hline
\hspace{1.5cm} {\bf DC-line} & \null \\ \hline
Series parameters per km ($R'_{dc,12}$/$L'_{dc,12}$) &0.0137 $\Omega$/km / 0.9339 mH/km \\
Shunt parameters per km ($C'_{cc,12}$) & 0.0119 $\mu$F/km  \\
Length ($\ell_{12}^{0}$) & 240 km \\
Total DC-bus eq. capacitance ($C_{dc,i}$) & 195 $\mu$F  \\
 ($C_{dc,i}=C_{VSC,i} + \sum_{j \neq i} C_{cc,ij}/2$) &  \\
\hline
\end{tabular}
\label{tab:sim1_system_parameters}
\end{center}
\end{table}

\section*{Acknowledgment}
Work supported by the Spanish Government under RETOS Project  Ref. RTI2018-098865-B-C31 and by Madrid Regional Government under PROMINT-CM Project  Ref. S2018/EMT-4366.


\begin{thebibliography}{10}
\providecommand{\url}[1]{#1}
\csname url@samestyle\endcsname
\providecommand{\newblock}{\relax}
\providecommand{\bibinfo}[2]{#2}
\providecommand{\BIBentrySTDinterwordspacing}{\spaceskip=0pt\relax}
\providecommand{\BIBentryALTinterwordstretchfactor}{4}
\providecommand{\BIBentryALTinterwordspacing}{\spaceskip=\fontdimen2\font plus
\BIBentryALTinterwordstretchfactor\fontdimen3\font minus
  \fontdimen4\font\relax}
\providecommand{\BIBforeignlanguage}[2]{{%
\expandafter\ifx\csname l@#1\endcsname\relax
\typeout{** WARNING: IEEEtran.bst: No hyphenation pattern has been}%
\typeout{** loaded for the language `#1'. Using the pattern for}%
\typeout{** the default language instead.}%
\else
\language=\csname l@#1\endcsname
\fi
#2}}
\providecommand{\BIBdecl}{\relax}
\BIBdecl

\bibitem{Flourentzou2009}
N.~Flourentzou, V.~G. Agelidis, and G.~D. Demetriades, ``{VSC-Based HVDC Power
  Transmission Systems: An Overview},'' \emph{IEEE Transactions on Power
  Electronics}, vol.~24, no.~3, pp. 592--602, 2009.

\bibitem{Inelfe2016}
J.~Bola, R.~Rivas, R.~Fern\'{a}ndez-Alonso, G.~P\'{e}rez, J.~Hidalgo, L.~M.
  Coronado, C.~Long\'{a}s, S.~Sanz, G.~Lemarchand, J.~Roguin, and D.~Glaise,
  ``{Operational experience of new Spain-France HVDC interconnection},'' in
  \emph{Proc. CIGRE Session, paper B4-117, Paris, France}, 2016, pp. 1--13.

\bibitem{Michi2019}
L.~Michi, E.~M. Carlini, T.~B. Scirocco, G.~Bruno, R.~Gnudi, G.~Pecoraro, and
  C.~Pisani, ``{AC Transmission Emulation Control Strategy in VSC-HVDC systems:
  general criteria for optimal tuning of control system},'' in \emph{Proc. AEIT
  HVDC International Conference, Florence, Italy}, 2019, pp. 1--6.

\bibitem{Marz2014}
M.~Marz, K.~Copp, A.~Manty, D.~Dickmander, J.~Danielsson, F.~Johansson,
  P.~Holmberg, P.-E. Bjorklund, H.~Duchen, P.~Lundberg, G.~Irwin, and
  S.~Sankar, ``{Mackinac HVDC Converter Automatic runback utilizing locally
  measured quantities},'' in \emph{Proc. CIGRE Conference, Toronto, Ontario,
  Canada}, 2014, pp. 1--9.

\bibitem{Danielsson2015}
J.~Danielsson, S.~Patel, J.~Pan, and R.~Nuqui, ``{Transmission Grid
  reinforcement with Embedded VSC-HVDC},'' in \emph{Proc. CIGRE US National
  Committee 2015 - Grid of the Future Symposium, Chicago, USA}, 2015, pp. 1--7.

\bibitem{Michi2020}
L.~Michi, E.~M. Carlini, G.~M. Giannuzzi, R.~Zaottini, C.~Pisani, F.~Allella,
  G.~Bruno, R.~Gnudi, A.~Pascucci, and G.~Pecoraro, ``{Dynamic stability issues
  of VSC-HVDC systems in AC Transmission Emulation Control: the Piossasco -
  Grande Ile case},'' in \emph{Proc. CIGRE Session, paper B4-118, Paris,
  France}, 2020, pp. 1--10.

\bibitem{IIT_ACemul2019}
J.~Renedo, L.~Rouco, L.~Sigrist, and A.~Garc\'ia-Cerrada, ``{Impact of
  AC-line-emulation controllers in VSC-HVDC link on inter-area-oscillation
  damping},'' in \emph{Proc. 45th Annual Conference of the IEEE Industrial
  Electronics Society (IECON), Lisbon, Portugal}, 2019, pp. 1--6.

\bibitem{Inelfe2020}
A.~D\'iaz, G.~Torresan, A.~Cord\'on, L.~Coronado, S.~Sanz, J.~Peir\'o,
  A.~Hern\'andez, J.~P\'erez, F.~Xavier, C.~Cardozo, and S.~Akkari,
  ``{Improvement of the oscillatory behaviour of the HVDC link between Spain
  and France},'' in \emph{Proc. CIGRE Session, paper B4-130, Paris, France},
  2020, pp. 1--10.

\bibitem{Kundur2004}
P.~Kundur, J.~Paserba, V.~Ajjarapu, G.~Andersson, A.~Bose, C.~Canizares,
  N.~Hatziargyriou, D.~Hill, A.~Stankovic, C.~Taylor, T.~{Van Cutsem}, and
  V.~Vittal, ``{Definition and classification of power system stability
  IEEE/CIGRE joint task force on stability terms and definitions},'' \emph{IEEE
  Transactions on Power Systems}, vol.~19, no.~3, pp. 1387--1401, 2004.

\bibitem{Johansson2004}
S.~G. Johansson, G.~Asplund, E.~Jansson, and R.~Rudervall, ``{Power system
  stability benefits with VSC DC-transmission systems},'' in \emph{Proc. CIGRE
  Session, Paris, France}, 2004, pp. 1--8.

\bibitem{Latorre2008}
H.~Latorre, M.~Ghandhari, and L.~S\"{o}der, ``{Active and reactive power
  control of a VSC-HVdc},'' \emph{Electric Power Systems Research}, vol.~78,
  no.~10, pp. 1756--1763, 2008.

\bibitem{Lukas2015}
L.~Sigrist, F.~Echavarren, L.~Rouco, and P.~Panciatici, ``{A fundamental study
  on the impact of HVDC lines on transient stability of power systems},'' in
  \emph{Proc. IEEE/PES PowerTech Conference, Eindhoven, Netherlands}, 2015, pp.
  1--6.

\bibitem{Machowski2013}
J.~Machowski, P.~Kacejko, L.~Nogal, and M.~Wancerz, ``{Power system stability
  enhancement by WAMS-based supplementary control of multi-terminal HVDC
  networks},'' \emph{Control Engineering Practice}, vol.~21, no.~5, pp.
  583--592, 2013.

\bibitem{Eriksson2014a}
R.~Eriksson, ``{Coordinated Control of Multiterminal DC Grid Power Injections
  for Improved Rotor-Angle Stability Based on Lyapunov Theory},'' \emph{IEEE
  Transactions on Power Delivery}, vol.~29, no.~4, pp. 1789--1797, 2014.

\bibitem{Fuchs2014}
A.~Fuchs, M.~Imhof, T.~Demiray, and M.~Morari, ``{Stabilization of Large Power
  Systems Using VSC - HVDC and Model Predictive Control},'' \emph{IEEE
  Transactions on Power Delivery}, vol.~29, no.~1, pp. 480--488, 2014.

\bibitem{Tang2016}
G.~Tang, Z.~Xu, H.~Dong, and Q.~Xu, ``{Sliding Mode Robust Control Based
  Active-Power Modulation of Multi-Terminal HVDC Transmissions},'' \emph{IEEE
  Transactions on Power Systems}, vol.~31, no.~2, pp. 1614--1623, 2016.

\bibitem{iitcontrolQ2017}
J.~Renedo, A.~Garc\'{i}a-Cerrada, and L.~Rouco, ``{Reactive-Power Coordination
  in VSC-HVDC Multi-Terminal Systems for Transient Stability Improvement},''
  \emph{IEEE Transactions on Power Systems}, vol.~32, no.~5, pp. 3758--3767,
  2017.

\bibitem{Fan2018}
X.~Fan, J.~Shu, and B.~Zhang, ``{Coordinated Control of DC Grid and Offshore
  Wind Farms to Improve Rotor-Angle Stability},'' \emph{IEEE Transactions on
  Power Systems}, vol.~33, no.~4, pp. 4625--4623, 2018.

\bibitem{iitcontrolQ_local2019}
J.~Renedo, L.~Rouco, A.~Garc\'{i}a-Cerrada, and L.~Sigrist, ``{A
  communication-free reactive-power control strategy in VSC-HVDC multi-terminal
  systems to improve transient stability},'' \emph{Electric Power Systems
  Research}, vol. 174, no. 105854, pp. 1--13, 2019.

\bibitem{JuanCarlos_transient_stab2020}
J.~C. Gonzalez-Torres, G.~Damm, V.~Costan, A.~Benchaib, and
  L.~Lamnabhi-Lagarrigue, ``{A novel distributed supplementary control of
  Multi-Terminal VSC-HVDC grids for rotor angle stability enhancement of AC/DC
  systems},'' \emph{IEEE Transactions on Power Systems}, vol. doi:
  10.1109/TPWRS.2020.3030538, pp. 1--12, 2020.

\bibitem{Cole2010}
S.~Cole, J.~Beerten, and R.~Belmans, ``{Generalized Dynamic VSC MTDC Model for
  Power System Stability Studies},'' \emph{IEEE Transactions on Power Systems},
  vol.~25, no.~3, pp. 1655--1662, 2010.

\bibitem{Beerten2014}
J.~Beerten, S.~Cole, and R.~Belmans, ``{Modeling of Multi-Terminal VSC HVDC
  Systems With Distributed DC Voltage Control},'' \emph{IEEE Transactions on
  Power Systems}, vol.~29, no.~1, pp. 34--42, 2014.

\bibitem{jrenedoPSSE2017}
J.~Renedo, A.~Garc\'{i}a-Cerrada, L.~Rouco, L.~Sigrist, I.~Egido, and S.~{Sanz
  Verdugo}, ``{Development of a PSS/E tool for power-flow calculation and
  dynamic simulation of VSC-HVDC multi-terminal systems},'' in \emph{Proc. 13th
  IET International Conference on AC and DC Power Transmission, Manchester,
  UK}, 2017, pp. 1--6.

\bibitem{Daelemans2009}
G.~Daelemans, K.~Srivastava, M.~Reza, S.~Cole, and R.~Belmans, ``{Minimization
  of steady-state losses in meshed networks using VSC HVDC},'' in \emph{Proc.
  IEEE/PES General Meeting, Calgary, AB, Canada}, 2009, pp. 1--5.

\bibitem{Beerten2012}
J.~Beerten, S.~Cole, and R.~Belmans, ``{Generalized Steady-State VSC MTDC Model
  for Sequential AC/DC Power Flow Algorithms},'' \emph{IEEE Transactions on
  Power Systems}, vol.~27, no.~2, pp. 821--829, 2012.

\bibitem{Kundur1994}
P.~Kundur, \emph{Power System Stability and Control}.\hskip 1em plus 0.5em
  minus 0.4em\relax McGraw Hill Education, 1994.

\bibitem{iit_coord_control_delay2019}
J.~Renedo, A.~Garc\'{i}a-Cerrada, L.~Rouco, and L.~Sigrist, ``{Coordinated
  Control in VSC-HVDC Multi-Terminal Systems to Improve Transient Stability:
  The Impact of Communication Latency},'' \emph{Energies}, vol.~12, pp. 1--32,
  2019.

\end{thebibliography}


\end{document}